# An application of bole surface growth model: a transitional status of '-3/2' rule

Gavrikov V.L.

**Introduction**

The famous '-3/2' thinning rule appeared in 1960-ties as a result of studies of the relationships between density of individuals and their mean volume/weight (in logarithmic representations) in plant populations (Yoda et al., 1963). For the sake of justice, one has to note that the idea that forest stand density and mean sizes of plants in the population are self-limiting has a much deeper story. A hundred years ago, forest practitioners had known that the number of trees in a forest stand is inversely related to the mean sizes of trees (see, e.g. Frothingham, 1914). Reineke (1933) elaborated the understanding mathematically to find a relationship between stand density and mean quadratic diameter at breast height (dbh) called Reineke's rule and to formulate his stand density index. Some authors (Perala et al., 1999) explicitly give preference to Reineke's rule before '-3/2' rule arguing the latter to be purely descriptive. Hilmi (1955) explored the same hypothesis as Yoda et al. (1963) did – geometric similarity – but studied another relationship – between linear dimensions of trees and the stand density and got, naturally, the other slope of self-thinning curves (–1/2).

Nevertheless, it is the '-3/2' rule that received the abundant response in publications, it had been also understood to be an important empirical generalization (Kofman, 1986). During the decades followed the first publications by the authors of the '-3/2' rule a substantial amount of data have been published, both in support to the rule and in the opposition to it. Kofman (1986) expressed an opinion that the deviation from the '-3/2' rule in published data were due to violation of the primary supposition made to derive the rule, i.e. the hypothesis of the geometrical similarity. A few years later, Newton and Smith (1990) gave an illustrative example to this opinion. Comparing various data on self-thinning in black spruce they received two distinct groups of data. Some data were consistent with the thinning rule, while others had sufficiently deviating parameters from the expected ones. The authors concluded that allometric relationships including diameters of trees were at least partly responsible for these deviations.

Besides the comparison of the thinning rule with empirical data, a number of efforts have been done to rethink the role and the status of the '-3/2' rule. Lonsdale (1990) came to a conclusion that by the time of his publication no evidences of the rule holding were available. He nevertheless admitted that the existence of an ideal limiting line could not be rejected until crucial experiments were performed. Hamilton et al. (1995) commented that the size-density trajectories followed by plant populations under self-thinning did not necessarily have a –3/2 slope. However, in their view, a more general assertion of some power rule giving an upper limit to self-thinning populations remained intact. West et al. (1997) developed a theoretical framework within which a suggestion has been developed that the slope of the self-thinning should be –4/3 rather than –3/2. Usoltsev (2003) in a review of the discussions around '–3/2' rule noted that the edge of the criticism of the rule was directed at whether the slope of the self-thinning curve equaled to –3/2 or not. At the same time, according to him, no doubts have been expressed of the very existence of a line limiting the self-thinning.

To summarize these several viewpoint, in forty years past the publications on '–3/2' rule the scientific discussion tended to give the rule an asymptotic status. It means the admittance of a couple of theses: i) the slopes of thinning trajectories may deviate from –3/2 value, that is, this value is not a marked out constant; ii) in the space 'size–density', a limiting line exists that serves as a 'goal' for population dynamics which this dynamics tends to and, when reached, follows along the line.

It is worth also mentioning that the question of the thinning rule may be explored in two meanings, which are however not contradictory to each other. First, the densities of many forest stands could be related *statically* against their mean volumes/masses of trees. Second, the density of one particular forest stand could be followed *in time* as related to its mean volume/mass of trees.

The aim of the study was to test the asymptotic status of the '–3/2' thinning rule in the second meaning, both theoretically and empirically.

**Methods**

To test the status of the '-3/2' rule a model of bole surface growth of a forest stand has been explored. Despite of the model's simplified (or one could even say 'oversimplified') structure the model has been shown to hold enough similarity with real forest stands to predict essential parameters of 'size–density' and 'size–size' relationships (Gavrikov, 2014). Visually, the model corresponds to an idealized population of even-sized cones that can be described by the following relationships

$$r \propto \sqrt{\frac{1}{N^{\gamma_1}}} \text{ or } \left(N^{-\frac{1}{2}\gamma_1}\right), \tag{1}$$

$$l \propto r^{\frac{2}{\gamma_2}-1}, \tag{2}$$

where $N$ is the quantity (or density) of figures, $r$ and $l$ stand for radius of the cone base and generatrix, respectively, $\gamma_1$ and $\gamma_2$ are parameters. To derive these relations it was supposed that i) the cones are narrow and high enough to accept that the generatrix $l$ is approximately equal to the cone height $h$, i.e. $l \approx h$ and ii) if the quantity of the cones drops their horizontal size grows accordingly.

Importantly, it has been shown that when the total bole surface $\hat{S}$ was independent of $N$, i.e. $\hat{S}(N) = C = const$, then $\gamma_1 = \gamma_2$. The latter equality means that the power exponent in (1) can be directly predicted from (2) and vise versa. If $\hat{S}$ depends on $N$, i.e. $\hat{S}(N)$ grows or drops with the decrease in $N$ (as a result of self-thinning) then relationships between $\gamma_1$ and $\gamma_2$ can be predicted.

The model provides an opportunity to formulate a simple and clear expression of the total bole surface area as

$$\hat{S} = \pi \cdot l \cdot r \cdot N. \tag{3}$$

The formula (3) along with (1) and (2) and the established properties of the model (Gavrikov, 2014) were used to explore the question about the status of the '–3/2' thinning rule.

To verify the results of the model analysis a number of data sources were used. The first group of data originates from levels-of-growing-stock studies in Douglas-fir plantations performed since 1960-ies by the Forest Service of the US. At my disposal, there were five datasets: Hoskins (Marshall, Curtis, 2001), Iron Creek (Curtis et al., 2009a), Rocky Brook (Curtis et. al., 2009b), Skykomish and Clemons (King et al., 2002) studies reports. The reports contain abundant and detailed information on the plantations stand development with many parameters like age of the plots, stand density $N$, mean height $H$, mean dbh $D$, quadratic mean diameter $Dq$, volume stock $Vs$, and others. The data on the development of control plots were of special interest for this study because they were not subjected to any treatment and therefore reflected a natural self-induced forest stand dynamics.

The total bole surface area BS of a forest stand was estimated through

$$BS = \frac{1}{2} \pi \cdot Dq \cdot H \cdot N \tag{3a}$$

that has been earlier found to be a useful approximation to stand surface area (Inoue, 2004). The use of quadratic mean diameter instead of mean dbh is of quite importance because in the forest science the volume stock is calculated via the quadratic mean diameter. Therefore the measure has been used to provide a compatibility with the data on the volume stock the analysis of which dynamics of one of the key points in the study.

The second group of data was extracted from the book by Usoltsev (2010) who gathered a large collection of data (altogether over 10,000 descriptions) on phytomass of various fractions in forests, mostly from European and Asian studies. The descriptions contain site quality indices from I (the best) to V (the worst) according to the Russian scale of bonitation, age of forest stand, its density *N*, mean height *H*, mean dbh *D*, and volume stock *Vs*. The descriptions are grouped by Usoltsev (2010) by the names of the authors and the following Scots pine (Pinus sylvestris) datasets were taken for the analysis:

1. Uspenskii, pine plantations in Tambov region (Russia), site quality Ia, I, and II
2. Mironenko, pine plantations in Tambov region (Russia), site quality Ia and I
3. Kozhevnikov, pine plantations (Belorussia), site quality I
4. Gruk, pine plantations (Belorussia), site quality I
5. Gabeev, natural pine forests at Novosibirsk (Russia), site quality I
6. Kurbanov, natural pine forests at Yoshkar-Ola (Russia), site quality I
7. Heinsdorf, natural pine forest at Eberswalde (Germany), site quality Ia, I, and II

It should be noted that the Scots pine data are static, which means that the data are descriptions of many forest stands at a particular age of the stands. At the same time, the model is dynamic, which means that it describes a single forest stand evolving in time (although the time is substituted here by fall of the stems density *N*). Therefore the data are suitable for testing of the model only to that extent to which the individual forest stands can be presented as a time series of a single forest stand dynamics. To meet the requirement I had to subsample from the above enumerated Scots pine datasets to extract a simple monotonous tendency in a forest stand dynamics: increase, constancy, or decrease of the total bole surface area. Such a separation of different tendencies is of importance for the analysis presented below. An advantage of the datasets is that one can find there descriptions of the forest stands that are old enough to show sufficient decrease in measures of production, which is a rare case for data like those of Douglas-fir as most of the direct experiments operate with younger and hence good growing forests. A consideration of the dynamics stages when forest stands experience decrease of production is also of quite an importance for the analysis.

The total bole surface area for the pine data was estimated through

$$BS = \frac{1}{2}\pi \cdot D \cdot H \cdot N .\qquad(3b)$$

It should be noted that the parameter of mean dbh is used in (3b) because this is the only opportunity as no data on quadratic mean diameter are contained in the database by Usoltsev (2010).

The estimations of regression parameters in the relationships studied were performed with the help of STATISICA 6 software.

**Results and discussion**

*Theoretical analysis*

Let us consider three types of relationships that describe the dynamics of an even-aged forest stand, these are dependencies of: i) mean radius of tree trunks *r* on the stand density, i.e. *r(N)*, ii) total bole surface area $\hat{S}$ on the stand density, i.e. $\hat{S}(N)$ and iii) mean per tree volume stock $\overline{Vs}$ on the stand density, i.e. $\overline{Vs}(N)$. The relationship *r(N)* according to the model takes the form (1) while $\hat{S}(N)$ and $\overline{Vs}(N)$ are supposed to also may take a form of a power function as follows:

$$\hat{S}(N) \propto N^{\alpha}, \qquad (4)$$

$$\overline{Vs}(N) \propto N^{\beta}. \qquad (5)$$

The relations (4) and (5) are not to say that the power function is the only or the best form to describe the relations. It is only implied that real non-monotonous curves may be divided into more or less monotonous segments within which a certain tendency can be seen, e.g., fast growth, slow growth, slow decrease etc. Supposedly, it is these segments that may be approximated by a power function and the exponent of the function gives an estimation of the intensity of the tendency, growth or fall. Another reason to use the form of a power function is that it allows one to easily study the relationships through a purely analytical approach.

One can note that the exponent β in (5) is the very factor that has been a keystone of the thinning rules analysis as it exactly corresponds to the slope of a self-thinning line in log–log coordinates. The interrelationships of $\gamma_1$, α and β are in the center of the research below.

According to (3) an expression for $\overline{Vs}(N)$ may be found. A multiplication of the both sides of the equality by *r* gives

$$r \cdot \hat{S} = \pi \cdot l \cdot r^2 \cdot N = \hat{Vs}$$

where $\hat{Vs}$ stands for the total volume stock of the model population. Dividing of the expression by *N* produces then a basic formula for the mean per tree volume stock as

$$\overline{Vs} = \frac{r \cdot \hat{S}}{N}. \qquad (6)$$

The expression (6) may be analyzed by various methods since the relationships *r(N)* and $\hat{S}(N)$ can be known or assessed. Let us take the relationships in the form (1) and (4), respectively. Then (6) can be rewritten as

$$\overline{Vs} = K \cdot N^{-\frac{\gamma_1}{2}} \cdot N^{\alpha} \cdot N^{-1} = K \cdot N^{\alpha - \frac{\gamma_1}{2} - 1} \qquad (7)$$

where *K* is a normalization constant. It follows from (7) straightforwardly that

$$\beta = \alpha - \frac{\gamma_1}{2} - 1, \qquad (8)$$

which permits a couple of inferences regarding the main question of the study. The first one sounds like *there cannot be one unique thinning line slope β that fits with every forest stand*. In fact, both $\gamma_1$ and α are highly likely to be very individual for particular forest stands subjected to self-thinning. The parameters should depend on plenty of internal and external factors as spatial distribution of trunks, genetic variety of the population etc. It is hard to imagine that $\gamma_1$ and α from different forest stands are so finely tuned to each other to give the same β in all cases.

Another inference from (8) is that *even within one particular forest stand the slope β does not remain constant in the course of time*. Indeed, there is no forbidding for both $\gamma_1$ and α varying during the life of individual forest stand. Especially it may concern the latter parameter as at growing stage of $\hat{S}(N)$ α < 0, when $\hat{S}(N)$ remain constant α ≈ 0 and when $\hat{S}(N)$ falls α > 0. Even if $\gamma_1$ alters correspondingly it is unlikely to expect $\gamma_1$ and α to exactly compensate each other. This can be easily demonstrated on the example of the most prolonged dataset on Douglas-fir growth, the Hoskins experiment (Marshall, Curtis, 2001) (fig. 1).

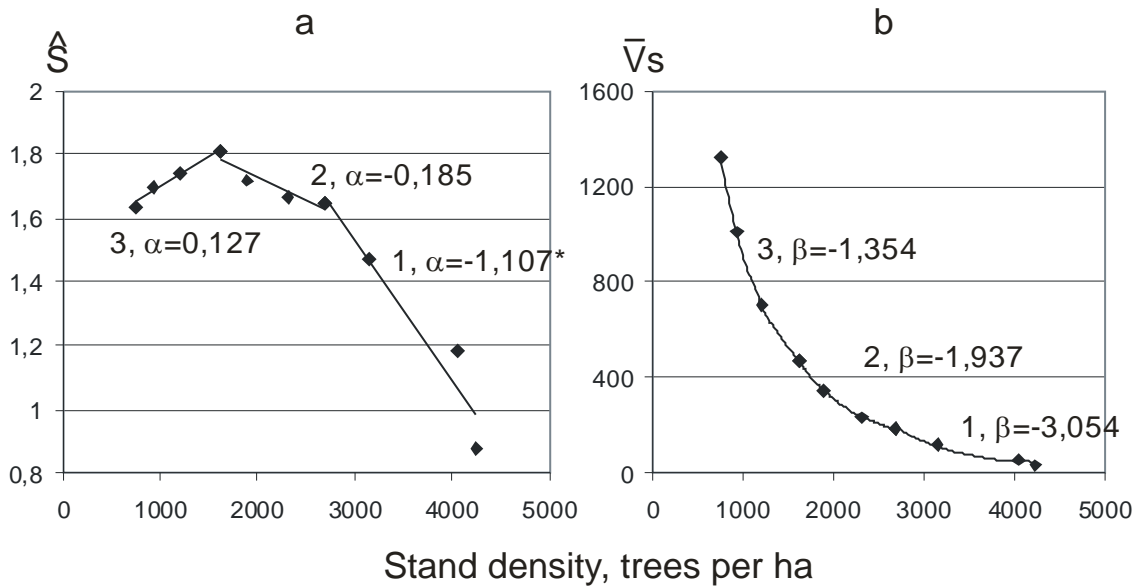

Fig. 1. Alterations of exponents in the relationships $\hat{S}(N) \propto N^\alpha$ (**a**) and $\overline{V}s(N) \propto N^\beta$ (**b**) in Hoskins experiment (Marshall, Curtis, 2001) in the course of growth of the same forest stand. The $\hat{S}$ axis denotes the estimation of the bole surface area according to (3a) in ha. The $\overline{V}s$ axis denotes the mean per tree trunk volume in cubic decimeters. The straight lines in (**a**) serve merely to mark the defined stages of dynamics, key: 1 – fast growth, 2 – slow growth, 3 – decrease of $\hat{S}$. The values of β in (**b**) were estimated for the same stages of the relationship $\overline{V}s$. With asterisk * the value is denoted that is significant at p < 0,1. All other estimated parameters

are significant at p < 0,05. The values of $\gamma_1$ of the relationship *r(N)* for the same defined stages are: 1 – 2,114; 2 – 1,277; 3 – 1,025 (all of them are significant at p < 0,05).

In the fig. 1a, the dynamics of the total bole surface area estimated through (3a) was visually divided into more or less monotonous segments (marked by straight lines) that corresponded to the defined stages of growth: fast growth, slow growth, and decrease. Then for the same stages in the relationship $\overline{V}s(N)$ the parameter β was estimated (fig. 1b). It is clear from the example that the slope in $\overline{V}s(N)$ evolves with the drop in the stand density, and more precisely, it evolves from the value β < –3/2 to β > –3/2, which will be dealt with in the later sections of the study.

*How good the model predicts the slope in $\overline{V}s(N)$*

In the above given example, the values of β were estimated through regression procedures, so the have so-to-say an empirical status. At the same time however it is worth comparing the regression values with those calculated with (8) to see how the formula (8) good in principle.

To evaluate the formula (8) all the datasets used were treated as follows. The curves of the total bole surface area dynamics $\hat{S}(N)$ were divided into more or less monotonous segments, analogically to the fig. 1a, with each segment expressing a definite tendency of growth. Then the parameters $\gamma_1$ and α were estimated for the segments and with the help of them the calculated $β_c$ were found through (8). For the same segments, the regression values of $β_r$ were found. A comparison of $β_r$ and $β_c$ is given in the tables 1 and 2 and in fig. 2.

Table 1. Estimated ($\gamma_1$, α, $β_r$) and calculated ($β_c$) parameters for the Douglas-fir database.

| Dataset ID | Tendency of segment[+] | $\gamma_1$[†] | α[†] | $β_r$[†] | $β_c$ | ($β_c$– $β_r$)/$β_c$, % |
|---|---|---|---|---|---|---|
| Hoskins | fast growth | 2,114 | -1,107* | -3,054 | -3,164 | 3,49% |
| Hoskins | slow growth | 1,277 | -0,185 | -1,937 | -1,823 | -6,22% |
| Hoskins | decrease | 1,025 | 0,127 | -1,354 | -1,385 | 2,24% |
| IronCreek | fast growth | 4,116 | -3,239 | -5,484 | -6,297 | 12,91% |
| IronCreek | slow growth | 1,210 | -0,122* | -1,720 | -1,727 | 0,41% |
| RockyBrook | fast growth | 4,555 | -3,945 | -6,731 | -7,223 | 6,81% |
| RockyBrook | slow growth | 1,320 | -0,237** | -1,761 | -1,897 | 7,15% |
| Skykomish | fast growth | 29,729** | -30,113* | -42,746* | -45,977 | 7,03% |
| Skykomish | slow growth | 2,016 | -1,017 | -2,895 | -3,026 | 4,32% |
| Clemons | fast growth | 12,531* | -12,081 | -18,074 | -19,346 | 6,57% |
| Clemons | slow growth | 1,640 | -0,634 | -2,358 | -2,454 | 3,90% |

[+]it implies the slope of monotonous segment in the $\hat{S}(N)$ curve

*significant at p < 0,1

Table 2. Estimated ($\gamma_1$, $\alpha$, $\beta_r$) and calculated ($\beta_c$) parameters for the Scots pine database.

| Dataset ID | Tendency of segment[+] | $\gamma_1$[†] | $\alpha$[†] | $\beta_r$[†] | $\beta_c$ | ($\beta_c - \beta_r$)/$\beta_c$, % |
|---|---|---|---|---|---|---|
| Uspenskii I | fast growth | 1,809 | -0,972 | -2,144 | -2,877 | 25,45% |
| Uspenskii I | flat | 1,103 | -0,005** | -1,425 | -1,556 | 8,40% |
| Uspenskii I | decrease | 1,038 | 0,118 | -1,333 | -1,401 | 4,80% |
| Uspenskii Ia | flat | 1,105 | -0,006** | -1,434 | -1,558 | 7,97% |
| Uspenskii Ia | decrease | 1,037 | 0,118 | -1,338 | -1,400 | 4,47% |
| Uspenskii II | decrease | 1,045 | 0,100 | -1,348 | -1,422 | 5,20% |
| Mironenko I | flat | 1,229 | -0,016** | -1,544 | -1,630 | 5,30% |
| Mironenko Ia | slow growth | 1,267 | -0,119* | -1,642 | -1,753 | 6,31% |
| Kozhevnikov | growth | 1,434 | -0,380 | -1,705 | -2,097 | 18,67% |
| Gruk | growth | 1,522 | -0,719 | -1,926 | -2,480 | 22,32% |
| Gabeev | decrease | 1,151 | 0,154** | -1,230 | -1,422 | 13,52% |
| Heinsdorf I | decrease | 1,017 | 0,182 | -1,221 | -1,326 | 7,95% |
| Heinsdorf Ia | decrease | 1,029 | 0,156 | -1,225 | -1,359 | 9,85% |
| Heinsdorf Ia | sharp decrease | 0,981 | 0,234 | -1,189 | -1,257 | 5,40% |
| Heinsdorf II | slow decrease | 1,138 | 0,043 | -1,429 | -1,526 | 6,35% |
| Heinsdorf II | sharp decrease | 0,993 | 0,221 | -1,197 | -1,275 | 6,09% |
| Kurbanov | decrease | 0,775 | 0,360 | -1,196 | -1,027 | -16,47% |

[+]it implies the slope of monotonous segment in the $\hat{S}(N)$ curve

*significant at p < 0,1

** not significant at p < 0,1

†unmarked values are significant at p < 0,05

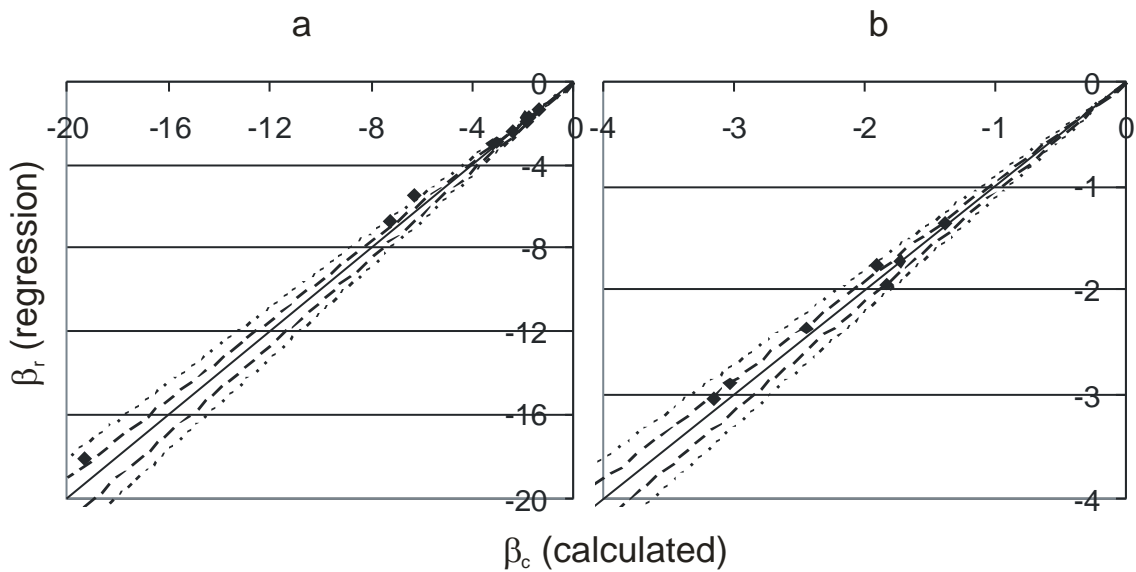

Fig. 2. Relation between calculated according to (8) ($\beta_c$) and estimated through regression ($\beta_r$) slopes of the relationship $\overline{V}s(N)$. **a**: the range of β shown up to –20, **b**: the upper right clump

of point is shown at a larger scale, the values of β are up to –4. The straight solid lines denote $\beta_c = \beta_r$, the dashed lines denote 5% and the dotted lines denote 10% deviations, respectively.

As follows from table 1 and fig. 2 there is a sufficiently good correspondence between the calculated from (8) $\beta_c$ and approximated directly from Douglas-fir data $\beta_r$. In most of the cases the model (i.e. $\beta_c$) slightly underestimates the values of $\beta_r$. Taking however into consideration that the model of forest stand structure is a very simple one suggests that such a result may be more than satisfactory.

On the other hand, a comparison of $\beta_c$ and $\beta_r$ for the Scots pine data reveals larger deviations between them (table 2). The most probable cause for the deviations is that the values of mean dbh, not those of quadratic mean diameter, were used in the calculations of the total bole surface area according to (3b). The quadratic mean diameter is known to be not only different from mean dbh but also related to it by a curvilinear relationship. This can be shown again on the example of Hoskins experiment data. The quadratic mean diameter $Dq$ measured consecutively over time in the Hoskins forest stand is related to the mean dbh $D$ as $Dq = 0{,}012D^2 + 0{,}04D + 4{,}47$ ($R^2 = 0{,}9994$), with the constant term being significant at $p < 0{,}05$. Such a result implies that when $D = 0$ $Dq \neq 0$, which may be a source of the deviations if the mean dbh is used instead of the quadratic mean diameter.

*Relation of $\beta_r$ to α in the context of -3/2 slope*

The model says (se e.g. (8)) that the value of α exerts a sufficient influence on the value of β. This is because, first, according to the observations α may be relatively big and, second, it may change its sign to the opposite. That is why it is worth looking at how estimated α relates to the estimated $\beta_r$.

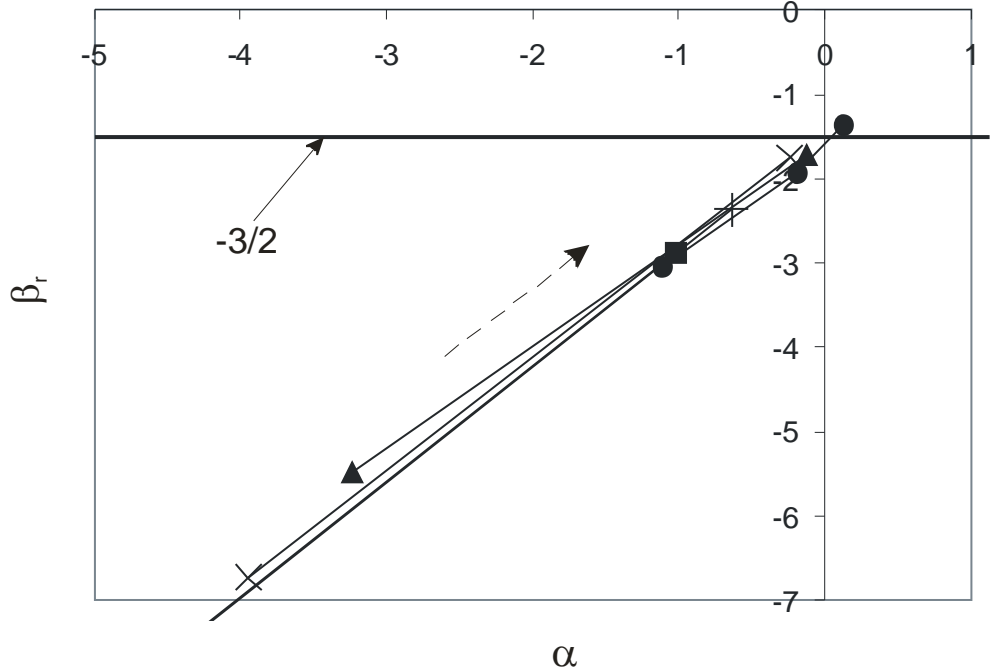

Fig. 3. Relationships between the self-thinning slope $\beta_r$ and the rate of the total bole surface area growth $\alpha$ for Douglas-fir database. Dataset key: ● – Hoskins, ▲ – Iron Creek, × – Rocky Brook, ■ – Skykomish, + – Clemons. Each symbol corresponds to a row in the table 1 (i.e. a segment of the growth curves). The solid lines connect symbols denoting the same dataset. The dashed arrow shows the direction of time dynamics. That is in terms of time, lower left symbols correspond to earlier moments while the upper right symbols correspond to later moments. The lower left points for Skykomish and Clemons are not shown as they lie far beyond the chart plane.

Fig. 3 summarizes the data on the relationships $\beta_r(\alpha)$ for the Douglas-fir database. A couple of findings can be derived from the chart. In spite of the substantial differences in the initial stand densities (4244 trees/ha in Hoskins vs. 1467 trees/ha in Skykomish experiments, see the correspondent references) and their different evolution over the time the trajectories of all the Douglas-fir stands go through a very narrow band in terms of the relationship $\beta_r(\alpha)$. This may imply that all the Douglas-fir plantations show an extremely high similarity of growth on the plane '$\alpha - \beta_r$'.

Another finding is that all the trajectories tend to approach the point of the chart where $\alpha = 0$ and $\beta_r = -3/2$. Unfortunately, only one Douglas-fir stand (Hoskins dataset) was sufficiently old to permit looking at what happens when the forest stand dynamics reaches the point. The case of Hoskins suggests that the point $\alpha = 0$, $\beta_r = -3/2$ is not an asymptote; the trajectory passes through it and continues to grow. This idea may be assessed with the help of the pine database as many of the pine forest stands have sufficiently high age and, more importantly, in terms of the total bole surface area show tendencies other that mere growth, e.g. a constant level or a decrease of the area.

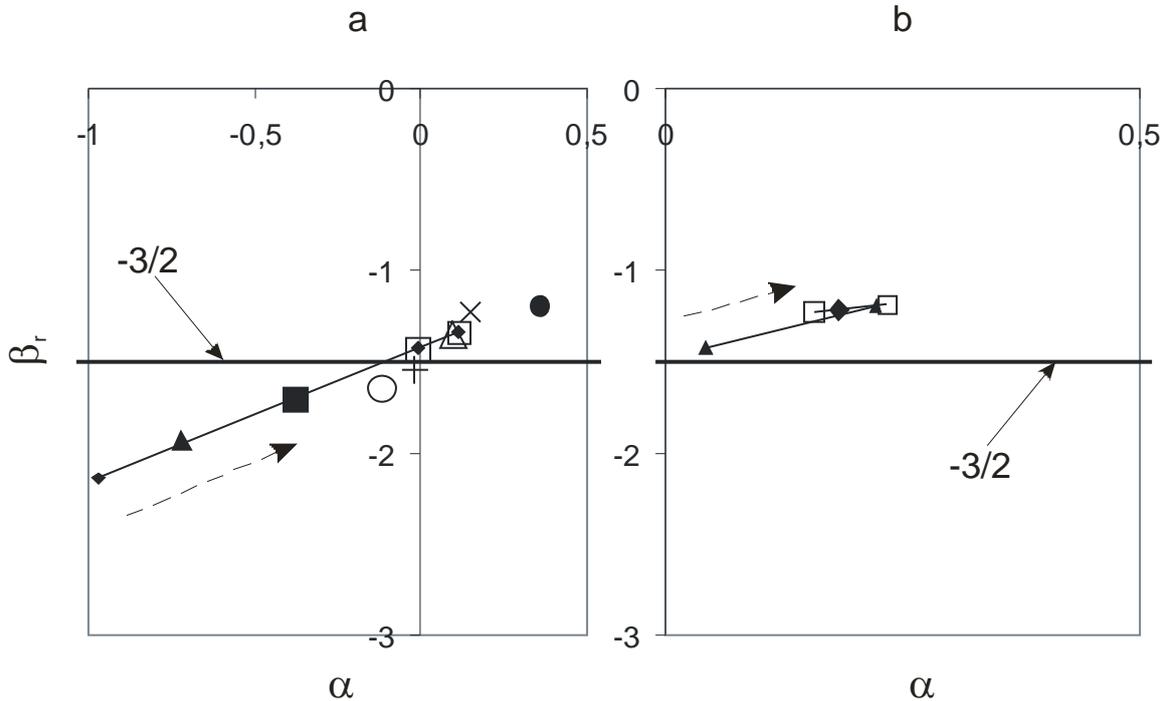

Fig. 4. Relationships between the self-thinning slope $\beta_r$ and the rate of the total bole surface area growth $\alpha$ for Scots pine database. **a**: ♦ – Uspenskii I, □ – Uspenskii Ia, △ – Uspenskii II, + – Mironenko I, ○ – Mironenko Ia, ■ – Kozhevnikov, ▲ – Gruk, × – Gabeev, ● – Kurbanov; **b**: ♦ – Heinsdorf I, □ – Heinsdorf Ia, ▲ – Heinsdorf II. Each symbol corresponds to a row in the table 2 (i.e. a segment of the growth curves). The solid lines connect symbols denoting the same dataset. The dashed arrows show the direction of time dynamics.

Details of the relationship $\beta_r(\alpha)$ for the Scots pine data are given in fig. 4. Some datasets in the figure are represented by a single point because they describe a single tendency in the $\hat{S}(N)$ evolution (see table 2). Nevertheless, both multipoint trajectories and single points representing quite different pine forest stands tend to lie in a narrow band and cross the point $\alpha = 0$, $\beta_r = -3/2$, as well as in the case of Douglas-fir. It is also clearly seen that the more positive values of $\alpha$ the farther the values of $\beta_r$ up from the line $-3/2$. The examples of Heinsdorf Ia and Heinsdorf II datasets (see table 2) show that the sharper decrease of $\hat{S}(N)$ the farther the value of $\beta_r$ shifts up from the line $-3/2$.

There are therefore two inferences that follow from the data. The first is that forest stands in the course of their self-thinning tend to cross the point $\alpha = 0$, $\beta_r = -3/2$ on the plane '$\alpha - \beta_r$'. Remember that $\alpha = 0$ means independence of the total bole surface area $\hat{S}$ of N, i.e. while the stand density proceeds to fall down the area $\hat{S}$ stays constant. Taking into account (8) shows that the only solution for $\alpha = 0$ and $\beta_r = -3/2$ at the same time is when $\gamma_1 = \gamma_2 = 1$. The latter equality and (2) imply that the height of trees is proportional to the radius of its base – that is that the geometrical similarity takes place over a certain span of the forest growth time. *This is the particular time when theses three peculiarities jump together: i) independence of the total bole surface area $\hat{S}$ of N, ii) geometrical similarity in the growth of trees, and hence iii) the value of the self-thinning slope $\beta_r = -3/2$.*

Another inference is *that the value $-3/2$ is not a goal (in the sense of an asymptote) for the evolution of the self-thinning curve slope*. The value of $\beta_r$ passes through the level of $-3/2$ and if $\alpha$ continues to grow then $\beta_r$ grows correspondingly.

Graphically, these ideas are presented in fig. 5.

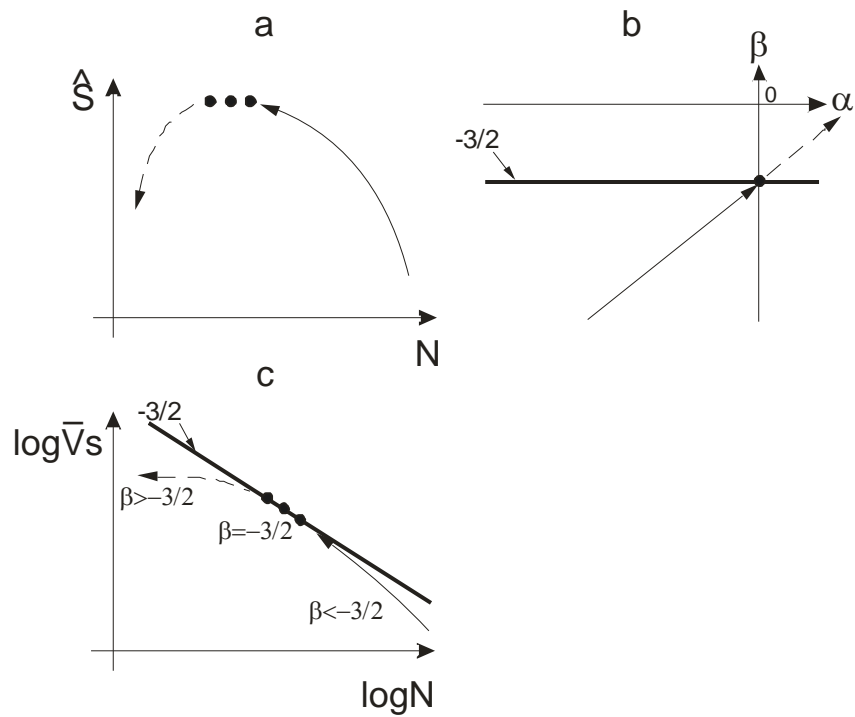

Fig. 5. Idealized descriptions of the main self-thinning processes: **a**: evolution of the total bole surface area with the stand density $\hat{S}(N)$; **b**: relationship between the power exponent $\alpha$ in $\hat{S}(N)$ and the power exponent $\beta$ in the self-thinning curve $\overline{V}s(N)$; **c**: the transition of $\beta$ from $\beta < -3/2$ through $\beta = -3/2$ to $\beta > -3/2$. Solid arrows denote the stage when $\beta < -3/2$ and the total bole surface area increases. Dots (●) designate the stage when $\beta = -3/2$ and the area stays constant. The dashed arrows indicate the stage when $\beta > -3/2$ and the area decreases.

**Conclusion**

In the research scope of forest stand self-thinning, the analysis performed above reveals a broad picture in which the '–3/2' rule takes a definite and special place. The application of the simple geometrical model to the Douglas-fir and Scots pine data suggests that the slope of the self-thinning curve $\overline{V}s(N)$ will not remain constant during the course of growth and self-thinning of a single forest stand. Most probable, at the initial stages of stand growth the slope will be less than –3/2 and at old ages of the stand the slope will be higher than –3/2. Inevitably, a time will come when the slope is exactly equals –3/2. In other words, the slope –3/2 is an obligatory state in the course of self-thinning of a forest stand.

At the very time of –3/2 slope two particular features coincide with it. One is that the total bole surface area remains constant. The length of the constancy stage would probably vary with species, initial stand densities, their spatial arrangements, conditions of growth, and other specific factors. Another feature of the time, from the point of view of the model, is that the model parameters $\gamma_1$ and $\gamma_2$ are equal to each other and unity, $\gamma_1 = \gamma_2 = 1$. The latter implies a geometric similarity in the growth of the forest stand, which is not in a contradiction with the '–3/2' rule as it had been formulated by its authors.

To put it shortly, the slope –3/2:

i) is a very specific and obligatory state in the process of forest stand growth and

ii) is not an asymptote but rather a transitional point (span) in the time of growth.

These two assertions may be called a 'transitional status of the '-3/2' rule'.

The geometric model of a forest stand (Gavrikov, 2014) has proved to be rather helpful at analyzing of real forest stand structure and dynamics. Despite of its extreme simplicity (it uses cones as representations of trees) the model looks like having enough similarity with real even-aged forests since the model's predictions are often reasonably close to measured values of power exponents.